\documentstyle[preprint,prc,aps]{revtex}
\begin{document}
\draft
\title{Theory of parity violation in compound nuclear states;\\
one particle aspects}
\author{N. Auerbach and V. Spevak}
\address{School of Physics and Astronomy, Tel Aviv University,
Tel Aviv 69978, Israel}
\date{\today}
\maketitle
\begin{abstract}
In this work we formulate the reaction theory of parity violation in
compound nuclear states using Feshbach's projection operator
formalism. We derive in this framework a complete set of terms that
contribute to the longitudinal asymmetry measured in experiments with
polarized epithermal neutrons. We also discuss the parity violating
spreading width resulting from this formalism. We then use the above
formalism to derive expressions which hold in the case when the
doorway state approximation is introduced. In applying the theory  we
limit ourselves in this work to the case when the parity violating
(PV) potential and the strong interaction are one-body. In this
approximation, using as the doorway the giant spin-dipole resonance
and employing well known optical potentials and a time-reversal even,
parity odd one-body interaction we calculate or estimate the terms we
derived. Among others, we evaluate numerically a new ``direct'' term
and the channel coupling term first derived by Lewenkopf and
Weidenm\"uller, this in addition to the evaluation of the main
``compound'' term. In our calculations we explicitly orthogonalize the
continuum and bound wave functions. We find the effects of
orthogonalization to be very important. Our conclusion is that the
present one-body theory cannot explain the average longitudinal
asymmetry found in the recent polarized neutron experiments. We also
confirm the discrepancy, first pointed out by Auerbach and Bowman,
that emerges, between the calculated average asymmetry and the parity
violating spreading width, when distant doorways are used in the
theory.
\end{abstract}

\narrowtext

\section{INTRODUCTION}
\label{sec:intro}

The recent studies of parity violation in compound nuclear states and
the observation \cite{Alfimenko,Bowman1,Bowman2} of large longitudinal
asymmetries in the transmission of polarized epithermal neutrons has
lead to a renewed interest in both the parity violating nuclear
interaction and the relevant structure of the compound nucleus. A
considerable number of theoretical papers was published on this
subject in the last 3-4 years \cite
{Flamb,Zarezky,Gudkov,AuR92,Bow-John,Koonin,Hussein,Lew-Weid,Auer-Bow}.
Most of the theoretical effort was directed towards the understanding
of the observed enhancements in the asymmetries and towards an
understanding of sign correlations in these asymmetries detected in
the studies with $^{232}Th$ \cite{Bowman2}. The common basis of the
majority of these theories was the use of a one-body parity
non-conserving potential that induces parity mixing in the nucleus. In
spite of this common basis the various works used different
mathematical approaches and calculated a number of different terms
that contribute to the parity violating asymmetries. In the present
work, starting from a one-body PV potential, we shall develop a
consistent framework in which we attempt to calculate or estimate all
the contributions emerging from such an approach.

The recent theoretical work as well as the pioneering papers in this
field \cite{Flamb,Zarezky} deal with a new type of experiments that
were first performed in the early eighties \cite{Alfimenko} and then
expanded in the in the early nineties mainly at the Los Alamos Neutron
Scattering Center (LANSCE) \cite{Bowman1,Bowman2}. In these
experiments one observes resonance scattering to p-wave resonances.
Using polarized neutron beams and unpolarized medium and heavy mass
targets one measures the longitudinal asymmetries:
\begin{equation}
P(E)=\frac{\sigma^{^{+}}(E)-\sigma^{^{-}}(E)}
{\sigma^{^{+}}(E)+\sigma^{^{-}}(E)} \ ,
\label{1}
\end{equation}
where $\sigma^{^{+}}$, $\sigma^{^{-}}$ are the total cross sections
for neutrons with positive and negative helicities respectively. A non
zero value for $P$ indicates that parity is not conserved in the
process. The large enhancements by factors $10^{5}$ predicted by
theory were indeed observed in the original experiments about ten
years ago \cite{Alfimenko} as well as in the more recent ones
\cite{Bowman1,Bowman2}. The great advantage of the recent improved Los
Alamos experiments is that one is now able to measure parity violation
of a number of $p_{1/2}$ resonances in the same nucleus. This enabled
one to apply statistical arguments in order to determine the mean
square PV matrix element and the {\em average\/} asymmetry $P$.

We should stress here that the excitation energies in the neutron plus
target nucleus  that are reached with eV neutrons are of the order of
\mbox{$5-10$ MeV}. In heavy nuclei this is the regime of the compound
nucleus where the density of states is very large, the spacing between
for example the \mbox{$J=\frac{1}{2}^{+}$} states is of the order of
\mbox{$10-20$ eV}.  This means that the wave function of each such
state contains about $10^{6}$ ``principal'' components, i.e.\
components that have comparable amplitudes each of the order
$10^{-3}$. The components are defined as many-body shell-model
configurations resulting from an appropriate single-particle basis.
Such large complexity of the compound nuclear states of course means
that statistical arguments should apply. It is therefore quite a
surprise that in the case of $^{232}Th$ the sign of the asymmetry $P$
measured for all the nine states, that show statistically significant
results, is positive. The average asymmetry for this nucleus was
determined \cite{Bowman2} to be about $8\%$.

The resonances, both the $s$ and $p$, are well separated, their width
being of the order of a very small fraction of an eV, while as
mentioned the separation is around 10 eV. Most of these resonances
were observed in the past in neutron scattering reactions. Higher $l$
resonances ($d,f$ etc.) are not seen because of the low
penetrabilities of such resonances. Among the $p$-wave resonances that
are observed, some have total spins of \mbox{$J=\frac{3}{2}$}. These
will not show up in the asymmetry measurements because parity
violation in such resonances occurs through the mixing with the
$d_{3/2}$ state and so the low penetrability will hinder the
observation of any asymmetry signal.

In the approach taken in this paper, as well as in some other recent
theoretical works we are able to incorporate aspects of relatively
simple nuclear structure and dynamics as well as statistical
behaviour. In the present work this duality will be quite explicit. In
some cases we shall average over certain class of compound states
while other states will be treated individually.

\section{ASYMMETRY $P$ AND THE $T$-MATRIX}
\label{sec:p-and-t}

The central quantity that is of interest in these parity violating
studies is the asymmetry given in Eq.\ (\ref{1}). In all the
experiments we discussed the asymmetry is measured on and off
resonance. The on resonance measurements are for the
\mbox{$J=\frac{1}{2}^{-}$} to which we shall refer as the $p_{1/2}$
resonances. (It should be clear that in  compound states the single
particle $p_{1/2}$ component is very small of the order of $10^{-5}$
admixture). For a $p_{1/2}$ resonance the asymmetry is:
\begin{equation}
P(E)=\frac{\sigma^{^{+}}_{p_{1/2}}(E)-\sigma^{^{-}}_{p_{1/2}}(E)}
{2\sigma^{0}_{p_{1/2}}(E)} \ ,
\label{2}
\end{equation}
where $\sigma^{^{+}}_{p_{1/2}}(E)$ and $\sigma^{^{-}}_{p_{1/2}}(E)$
are the helicity plus and minus on resonance total $p_{1/2}$ cross
sections, and $\sigma^{0}_{p_{1/2}}(E)$ is the total helicity
independent cross section for such resonance. In Appendix \ref{app:1}
we derive the relation between the asymmetry and the $T$-matrix. Here
we present this well known\cite{Gudkov} formula relating $P(E)$ to the
parity violating part of the $T$-matrix, $T^{PV}$:
\begin{equation}
P(E)=\frac{8\pi^{2}}{k^{2}}\frac{ImT^{PV}(E)}{\sigma^{0}_{p_{1/2}}(E)}
 \ .
\label{3}
\end{equation}

\section{REACTION THEORY}
\label{sec:react-theory}

We shall use the projection theory of Feshbach \cite{Feshbach} and its
extensions \cite{Feshbach-Lemmer,Auerbach72} to formulate the reaction
theory for the parity violating experiments in the compound nucleus.
Let us divide the complete space into the following subspaces relevant
to our problem. The $\{P\}$ space (the open channel space) will
contain the channels \mbox{$\phi_{p_{1/2}}\vert 0\rangle$} and
\mbox{$\phi_{s_{1/2}}\vert 0\rangle$}, where $\vert 0\rangle$ is the
ground state of the  target, $\phi_{p_{1/2}}$ and $\phi_{s_{1/2}}$ are
the continuum single-particle states. The corresponding projection
operator that projects onto this space will be denoted as $P$ (not to
be confused with $P(E)$ the asymmetry in Eq.\ (\ref{1})). The $\{Q\}$
space will contain all \mbox{$J^{\pi}=\frac{1}{2}^{+}$} bound states
$\vert q\rangle$ and the projection operator will be denoted as $Q$.
In the $\{R\}$ space we shall keep the
\mbox{$J^{\pi}=\frac{1}{2}^{-}$} bound states $\vert r\rangle$ and the
corresponding projection operator will be denoted by $R$. The rest of
the existing states will not be of interest to us and therefore are
not included in our considerations. As usual we shall assume that the
spaces
\mbox{$\{P, Q, R\}$} are mutually orthogonal,
\mbox{$P\cdot Q=$}\mbox{$P\cdot R=$}\mbox{$Q\cdot R=0$} and that
\mbox{$P^{2}=P$}, \mbox{$Q^{2}=Q$}, \mbox{$R^{2}=R$}. A state of total
spin \mbox{$J=\frac{1}{2}$} is then written in terms of the above
projection operators as:
\begin{equation}
\Psi=P\Psi+Q\Psi+R\Psi \ .
\label{4}
\end{equation}
Denoting by $H$ the full nuclear Hamiltonian which contains both
parity conserving and parity non-conserving parts, we can write the
following set of coupled equations:
\begin{eqnarray}
(E-H_{PP})P\Psi =H_{PR}R\Psi+H_{PQ}Q\Psi \ ,   \nonumber\\
(E-H_{RR})R\Psi =H_{RP}P\Psi+H_{RQ}Q\Psi \ ,  \label{5}  \\
(E-H_{QQ})Q\Psi =H_{QP}P\Psi+H_{QR}R\Psi \ ,  \nonumber
\end{eqnarray}
where the usual notation $H_{AB}$ stands for $AHB$. We shall now
eliminate the explicit reference to the $Q\Psi$ part of the wave
function by introducing the operator ${\cal H}$, given by:
\begin{equation}
{\cal H}=H+HQ\frac{1}{E-H_{QQ}}QH \ .
\label{6}
\end{equation}
Then we can write:
\begin{eqnarray}
(E-{\cal H}_{PP})P\Psi ={\cal H}_{PR}R\Psi \ , \nonumber\\
(E-{\cal H}_{RR})R\Psi ={\cal H}_{RP}P\Psi \ . \label{7}
\end{eqnarray}
The projection of the operator ${\cal H}$ in Eq.\ (\ref{6}) onto the
P-space ${\cal H}_{PP}$ we shall term as the optical model operator.
Solving formally one can write for $P\Psi$:
\begin{equation}
(E-{\cal H}_{PP}-{\cal H}_{PR}\frac{1}{E-{\cal H}_{RR}}{\cal H}_{RP})P\Psi=0
 \ .
\label{8}
\end{equation}
Let $\Phi^{(\pm)}_{c}$ be the solutions of
\mbox{$(E-{\cal H}_{PP})\Phi^{(\pm)}_{c}=0$}\ , i.e.
\begin{equation}
(E-(H+HQ\frac{1}{E-H_{QQ}}QH)_{PP})\Phi^{(\pm)}_{c}=0 \ .
\label{9}
\end{equation}

We note that in zero order of the PV interaction the optical model
operator contains for the $p$-wave channel only $H_{PP}$, while for
the $s$-wave the above operator contains also the more complicated
term which involves the $\vert q\rangle$-states. We should call the
attention to the fact that following the experimental approach we
attempt to calculate the various observables (cross section,
asymmetry) for individual $p$-wave resonances only. The coupled
equations (\ref{7}) will give rise to rapidly fluctuating scattering
amplitudes due to the $\{Q\}$ space. The averaging over this space
will be made at a later stage (see Sec.~\ref{sec:doorway}).

\subsection{Case1. No channel coupling.}
\label{subsec:nocoupling}

Let us first consider the case when the direct parity violating
coupling {\em between the channels\/} is neglected.
The Lippmann-Schwinger equation for $P\Psi^{(\pm)}_{c}$
(subscript $c$ denotes $p$ or $s$) can be written:
\begin{eqnarray}
P\Psi^{(\pm)}_{c}=&&\Phi^{(\pm)}_{c} \nonumber
\\&&+ \frac{1}{E^{^{(\pm)}}-{\cal H}_{PP}}{\cal H}_{PR}
\frac{1}{E^{^{(\pm)}}-{\cal H}_{RR}}{\cal H}_{RP}P\Psi^{(\pm)}_{c}
\nonumber \\
\label{10}
\end{eqnarray}
and more explicitly as:
\widetext
\begin{equation}
P\Psi^{(\pm)}_{c}=\Phi^{(\pm)}_{c}+ \frac{1}{E^{^{(\pm)}}-
{\cal H}_{PP}}{\cal H}_{PR}
\frac{1}{E^{^{(\pm)}}-{\cal H}_{RR}-{\cal H}_{RP}(E^{^{(\pm)}}-
{\cal H}_{PP})^{-1}{\cal H}_{PR}}{\cal H}_{RP}\Phi^{(\pm)}_{c} \ .
\label{11}
\end{equation}
\narrowtext
The $T$-matrix is:
\begin{eqnarray}
T_{sp}=&&T_{sp}^{(P)} \nonumber\\&&+
\sum_{r}\frac{\langle\Phi_{s}^{^{(-)}}\vert{\cal H}_{PR}\vert
r\rangle\langle r\vert{\cal H}_{RP}\vert\Phi_{p}^{^{(+)}}\rangle}
{E-\langle r\vert{\cal H}_{RR}\vert r\rangle-\langle r\vert
{\cal H}_{RP}G_{P}^{^{(+)}}{\cal H}_{PR}\vert r\rangle} \ , \nonumber
\\ \label{12}
\end{eqnarray}
where $\displaystyle G_{P}^{^{(\pm)}}=\frac{1}
{E^{^{(\pm)}}-{\cal H}_{PP}}$ and $T_{sp}^{(P)}$ is the non-resonant
part of the $T$-matrix due to ${\cal H}_{PP}$. From now on we shall
consider only the parity-violating part of the $T$-matrix:
\begin{equation}
T_{sp}^{PV}=T_{sp}^{(P) PV}+T_{sp}^{(res) PV} \ .
\label{13}
\end{equation}
Here $T_{sp}^{(res) PV}$ is the resonant part and $T_{sp}^{(P) PV}$ is
due to the parity-violating part of ${\cal H}_{PP}$:
\begin{equation}
T_{sp}^{(P) PV}=-\frac{1}{2\pi i}\langle\Phi_{s}^{^{(-)}}\vert
\Phi_{p}^{^{(+)}}\rangle \ ,
\label{14}
\end{equation}
where $\langle\Phi_{s}^{^{(-)}}\vert\Phi_{p}^{^{(+)}}\rangle$ is the
parity-violating $S$-matrix. Unless we take channel coupling into
account, $\Phi_{c}$ has a definite parity and the parity-violating
part of $T_{sp}^{(P)PV}$ is zero. To shorten our notation we write:
\begin{equation}
\langle r\vert{\cal H}_{RR}\vert r\rangle+\langle r\vert{\cal H}_{RP}
G_{P}^{^{(\pm)}}{\cal H}_{PR}\vert r\rangle=
E_{r}\mp\frac{i}{2}\Gamma_{r} \ .
\label{15}
\end{equation}

To first order in the parity violating interaction the second term in
Eq.\ (\ref{12}) gives two contributions. When $H_{RP}$ is parity
non-conserving we get the term:
\begin{equation}
T_{sp,dir}^{PV}=\sum_{r}\frac{\langle\Phi_{s}^{^{(-)}}\vert V^{PV}
\vert r\rangle\langle r\vert H_{RP}\vert\Phi_{p}^{^{(+)}}\rangle}
{E-E_{r}+\frac{i}{2}\Gamma_{r}} \ .
\label{16}
\end{equation}
Note that this term although resonant does not involve the mixing with
$\vert q\rangle$ states and therefore similarly to the case of
isobaric analog resonances \cite{Auerbach72} we will refer to it as
the ``direct'' term.

When the parity violation appears in \mbox{$\displaystyle
RHQ\frac{1}{E-H_{QQ}}QHP$} the corresponding part of the $T$-matrix,
called the compound term, will be:
\begin{equation}
T_{sp,comp}^{PV}=\sum_{r,q}\frac{\langle\Phi_{s}^{^{(-)}}\vert H_{PQ}
\vert q\rangle
\langle q\vert V^{PV}\vert r\rangle \langle r\vert H_{RP}\vert
\Phi_{p}^{^{(+)}}\rangle}{(E-E_{r}+\frac{i}{2}\Gamma_{r})(E-E_{q})}\ .
\label{17}
\end{equation}
We should comment here that the various compound states $\vert q\rangle$
have very small (but nonzero) decay widths and therefore the energies
$E_{q}$ do contain small imaginary parts preventing the expression in
Eq.\ (\ref{17}) and in all forthcoming equations involving
\mbox{$E-E_{q}$} in the denominator to became singular. For the sake
of simplicity of notation we have not included explicitly the above
imaginary part.

\subsection{Case2. Channel coupling.}
\label{subsec:coupling}

We consider now the case when there is direct  parity violating
coupling between the two channels that have opposite parity. We write
equation \mbox{$\displaystyle (E-{\cal H}_{PP})\Phi^{(\pm)}=0$} in the
form
\begin{equation}
\left (\begin{array}{c}{\cal H}_{PP}^{^{++}} \ {\cal H}_{PP}^{^{+-}}\\
{\cal H}_{PP}^{^{-+}} \ {\cal H}_{PP}^{^{--}} \end{array} \right )\left
(\begin{array}{c} \Phi_{s} \\ \Phi_{p} \end{array} \right )=E
\left (\begin{array}{c} \Phi_{s} \\ \Phi_{p} \end{array} \right )
\label{18}
\end{equation}
or
\begin{eqnarray*}
(E-&& H_{PP}^{^{++}}-H_{PQ}^{^{++}}\frac{1}{E-H_{QQ}}H_{QP}^{^{++}}-
H_{PP}^{^{--}}-H_{PP}^{^{-+}}\\
-&&H_{PQ}^{^{-+}}\frac{1}{E-H_{QQ}}H_{QP}^{^{++}}-H_{PP}^{^{+-}}-
H_{PQ}^{^{++}}\frac{1}{E-H_{QQ}}H_{QP}^{^{+-}})\\
\times (\Phi_{p}&&+\Phi_{s})=0\ ,
\end{eqnarray*}
where $\Phi_{p}$ and $\Phi_{s}$ are wave functions of $p_{1/2}$ and
$s_{1/2}$ channels. The superscripts denote parities. Parity-violating
terms of ${\cal H}_{PP}$ will give the $T_{sp}^{(P) PV}$ part of the
$T$-matrix:
\begin{eqnarray}
T_{sp}^{(P) PV}=&&\langle\Phi_{s}^{^{(-)}}\vert V^{PV}\vert
\Phi_{p}^{^{(+)}}\rangle \nonumber \\
&&+ \sum_{q}\frac{\langle\Phi_{s}^{^{(-)}}\vert H_{PQ}^{^{++}}\vert
q\rangle\langle q\vert V^{PV}\vert\Phi_{p}^{^{(+)}}\rangle}{E-E_{q}}
 \ .
\label{19}
\end{eqnarray}
To display the channel coupling contribution in the resonant part of
the $T$-matrix we write the solution $\hat{\Phi}$ for the $p-$wave
channel in first order of the PV-interaction:
\begin{equation}
\hat{\Phi}=\Phi_{p}+\frac{1}{E-H_{PP}}(H_{PP}^{^{-+}}+
H_{PQ}^{^{-+}}\frac{1}{E-H_{QQ}}H_{QP}^{^{++}})\Phi_{s} \ ,
\label{20}
\end{equation}
where $\Phi_{c}$ are solutions of the uncoupled equations with
definite parity, used in the previous section. Now inserting
$\hat{\Phi}^{^{(-)}}_{p}$ for $\Phi_{s}^{^{(-)}}$ in the resonant part
of the $T$-matrix in Eq.\ (\ref{12}) in the case, when ${\cal H}_{PR}$
and ${\cal H}_{RP}$ are parity-conserving operators, i.e.\ equal to
$H_{PR}$ and $H_{RP}$ we get in the first order of the PV-interaction
two terms, resulting from the {\em parity-admixed exit channels}:
\widetext
\begin{eqnarray}
&&T_{sp,(res,cc)}^{PV}=
\sum_{r}\frac{\langle\Phi_{s}^{^{(-)}}\vert V^{PV}\frac{1}{E^{(+)}-
H_{PP}}H_{PR}\vert r\rangle\langle r\vert H_{RP}\vert\Phi_{p}^{^{(+)}}
\rangle}{E-E_{r}+\frac{i}{2}\Gamma_{r}} \nonumber
\\&&+
\sum_{r,q}\frac{\langle\Phi_{s}^{^{(-)}}\vert H_{QP}\vert q\rangle
\langle q\vert V^{PV}\frac{1}{E^{(+)}-H_{PP}} H_{PR}\vert r\rangle
\langle r \vert H_{RP}\vert\Phi_{p}^{^{(+)}}\rangle}
{(E-E_{r}+\frac{i}{2}\Gamma_{r})(E-E_{q})}\ .
\label{21}
\end{eqnarray}
\narrowtext
Where:
\begin{equation}
\frac{1}{E^{(\pm)}-H_{PP}}=\sum_{c}\int dE'\vert\Phi_{c}(E')\rangle
\frac{1}{E^{(\pm)}-E'}\langle\Phi_{c}(E')\vert
\label{22}
\end{equation}
(subscript {\em cc\/} denotes channel coupling).

\section{EXPRESSIONS FOR THE ASYMMETRIES}
\label{sec:asymm}

To calculate the asymmetry $P(E)$ we make use of Eq.\ (\ref{3}) and
the equation for the total $p_{1/2}$ cross section given by:
\begin{equation}
\sigma^{0}_{p_{1/2}}\cong\frac{4\pi}{k^{2}}\frac{2\pi\vert
\langle r\vert H_{RP}\vert \Phi_{p}^{^{(+)}}\rangle\vert^{2}
\Gamma_{r}}{4(E-E_{r})^{2}+\Gamma_{r}^2}\ ,
\label{23}
\end{equation}
where $\Gamma_{r}$ is the total width of a $p_{1/2}$-wave resonance.
The statistical weight factor equals unity for $p_{1/2}$ wave
\cite{Bohr-Mottelson}. The on-resonance \mbox{$(E=E_{r})$} asymmetry
resulting from the direct term in Eq.\ (\ref{16}) is:
\begin{equation}
P_{dir}(E_{r})=-2\frac{Re(\langle\Phi_{s}^{^{(-)}}\vert V^{PV}
\vert r\rangle\langle r\vert H_{RP}\vert\Phi_{p}^{^{(+)}}\rangle)}
{\vert\langle r\vert H_{RP}\vert \Phi_{p}^{^{(+)}}\rangle\vert^{2}}\ .
\label{24}
\end{equation}
The on-resonance contribution of the compound part of the $T$-matrix
in Eq.\ (\ref{17}) to the asymmetry is:
\begin{eqnarray}
&&P_{comp}(E_{r}) \nonumber \\
&&=-2\sum_{q} \frac{Re(\langle\Phi_{s}^{^{(-)}}\vert H_{PQ}
\vert q\rangle
\langle q\vert V^{PV}\vert r\rangle \langle r\vert H_{RP}\vert
\Phi_{p}^{^{(+)}}\rangle )}
{\vert\langle r\vert H_{RP}\vert \Phi_{p}^{^{(+)}}\rangle\vert^{2}\cdot
(E_{r}-E_{q})} \ . \nonumber
\\ \label{25}
\end{eqnarray}
The asymmetry resulting from the channel coupling contribution in the
first term of Eq.\ (\ref{21}) is analogous to the direct term with the
matrix element
\mbox{$\displaystyle\langle\Phi_{s}^{^{(-)}}\vert V^{PV}
\frac{1}{E^{(+)}-H_{PP}}H_{PR}\vert r\rangle $}
replacing \mbox{$\langle\Phi_{s}^{^{(-)}}\vert V^{PV}\vert r\rangle $}
and given by:
\begin{eqnarray}
&&P_{cc}(E_{r})=-2\frac{1}{\vert\langle r\vert H_{RP}\vert
\Phi_{p}^{^{(+)}}\rangle\vert^{2}}Re\biggl
\{\langle r\vert H_{RP}\vert \Phi_{p}^{^{(+)}}(E_{r})\rangle
\nonumber \\ && \times
P.V.\int dE'\frac{
\langle\Phi_{s}^{^{(-)}}(E_{r})\vert V^{PV}\vert\Phi_{p}(E')\rangle
\langle \Phi_{p}(E')\vert H_{PR}\vert r\rangle}
{E_{r}-E'}\biggr\} \ . \nonumber \\
\label{26}
\end{eqnarray}
(The symbol $P.V.$ stands for ``principal value''). This contribution
was suggested by Lewenkopf and Weidenm\"uller \cite{Lew-Weid} as a
possible source of enhancement in $P_{cc}(E)$. The second term of the
$T$-matrix in Eq.\ (\ref{21}) gives the following contribution:
\begin{eqnarray}
&&P_{cc}(E_{r})=-2\frac{1}{\vert\langle r\vert H_{RP}\vert
\Phi_{p}^{^{(+)}}\rangle\vert^{2}} \nonumber \\
&&\times\sum_{q} Re\biggl\{\frac{
\langle\Phi_{s}^{^{(+)}}\vert H_{PQ}\vert q\rangle
\langle r\vert H_{RP}\vert\Phi_{p}^{^{(-)}}(E_{r})\rangle}{E_{r}-E_{q}}
\nonumber \\
&&\times
P.V.\int dE'\frac{\langle q\vert V^{PV}\vert\Phi_{p}(E')\rangle
\langle\Phi_{p}(E')\vert H_{PR}\vert r\rangle}{E_{r}-E'}\biggr\} \ .
\label{27}
\end{eqnarray}
Finally we shall write the non-resonant channel coupling contributions
to $P(E)$. Because we consider here only the $p_{1/2}$-resonance case
we shall evaluate the non-resonant contribution at the energy
\mbox{$E\approx E_{r}$} meaning that we can neglect the contribution
of the tails of $s$-wave resonances. (In the off-resonance region
$\sigma^{0}$ has normally a contribution from the tail of the nearest
$s$-resonance, which is usually bigger than the $p$-wave off-resonance
cross section). The first term of the non-resonant channel coupling
part of the $T$-matrix Eq.\ (\ref{19}) for $E=E_{r}$ gives:
\begin{equation}
P_{cc}(E_{r})=Im(
\langle\Phi_{s}^{^{(-)}}\vert V^{PV}\vert\Phi_{p}^{^{(+)}}\rangle)
\frac{\Gamma_{r}}
{\vert\langle r\vert H_{RP}\vert\Phi_{p}^{^{(+)}}\rangle\vert^{2}} \ .
\label{29}
\end{equation}
This term is equal to zero, if the the radial parts of $\phi_{p}$ and
$\phi_{s}$ functions are purely real.
The second term in Eq.\ (\ref{19}) yields for $E=E_{r}$:
\begin{eqnarray}
P_{cc}(E_{r})=&&\frac{\Gamma_{r}}
{\vert\langle r\vert H_{RP}\vert\Phi_{p}^{^{(+)}}\rangle\vert^{2}}
\nonumber \\
&&\times \sum_{q}\frac{Im(\langle\Phi_{s}^{^{(-)}}\vert
H_{PQ}^{^{++}}\vert q\rangle \langle q\vert V^{PV}\vert
\Phi_{p}^{^{(+)}}\rangle)}{E_{r}-E_{q}} \ .
\label{31}
\end{eqnarray}

We should stress here that a number of contributions we have written
here will be numerically small and not have an impact on the final
results. The numerical calculations and estimates of these various
contributions to \mbox{$P(E_{r})$} will be discussed in
Sec.~\ref{sec:numcalc}.

One can write all the expressions for the asymmetry in a concise form:
\begin{equation}
P(E_{r})=-2\frac{Re(\gamma_{r}^{\uparrow}\gamma_{r}^{(PV)\uparrow})}
{\vert\gamma_{r}^{\uparrow}\vert^{2}} \ ,
\label{32}
\end{equation}
where $\gamma_{r}^{\uparrow}$ is the neutron escape amplitude:
\begin{equation}
\gamma_{r}^{\uparrow}=\langle r\vert H_{RP}\vert
\Phi_{p}^{^{(+)}}\rangle
\label{33}
\end{equation}
and $\gamma_{r}^{(PV)\uparrow}$ is the {\em parity violating\/} escape
amplitude relevant to each of the processes discussed above. For
example in the direct process, described by Eq.\ (\ref{16}) the
corresponding $\gamma_{r}^{(PV)\uparrow}$ is:
\begin{equation}
\gamma_{r,dir}^{(PV)\uparrow}=\langle \Phi_{s}^{^{(-)}}\vert V^{PV}
\vert r\rangle \ ,
\label{34}
\end{equation}
while for the compound process in Eq.\ (\ref{17})
\begin{equation}
\gamma_{r,comp}^{(PV)\uparrow}=\sum_{q}\frac
{\langle \Phi_{s}^{^{(-)}}\vert H_{PQ}\vert q\rangle\langle q\vert
V^{PV}\vert r\rangle}{E_{r}-E_{q}} \ .
\label{35}
\end{equation}

\section{DOORWAY STATE APPROACH}
\label{sec:doorway}

The expressions for the asymmetry $P_{comp}(E)$ in Eq.\ (\ref{25}) are
very complicated because they involve in principle an infinite sum
over all compound states $\vert q\rangle$. This complexity is of
course inherent in the physics of the compound nucleus, and usually
statistical methods are adopted in order to calculate the observables
\cite{Gudkov,French}. However, it was recognized in the past that due
to the absence of strong multi-particle forces in the nuclear
Hamiltonian one can often reduce  the complexity of the problem by
introducing the notion of doorways \cite{Feshbach-Lemmer,Auerbach72}.
The idea is to select in the $\{Q\}$ space states which have direct
couplings via a one or two-body force either to the channel space or
the states in the $\{R\}$ space. These special selected states are the
doorways. Formally, we divide the $\{Q\}$ space into the subspace of
doorways denoted as $\{D\}$, the states belonging to it as $\vert
d\rangle$, and the corresponding projection operator as $D$. The rest
of the $\{Q\}$ space will be denoted as $\{Q'\}$ and its states as
\mbox{$\vert q'\rangle$}.
\begin{equation}
\{Q\}=\{D\}+\{Q'\}
\label{36}
\end{equation}
The  states $\vert d\rangle$ are not eigenstates even of $H_{QQ}$.
The doorway states thus are model states that are treated explicitly.
The influence of the $\vert q'\rangle$ is taken into account
indirectly by introducing new parameters into the theory, such as
widths, energy shifts, etc. which are usually treated
phenomenologically. For example in our discussion of Eq.\ (\ref{25})
and the PV matrix element \mbox{$\langle r\vert V^{PV}\vert q\rangle$}
we replace the $\vert q\rangle$ states by some doorways
\mbox{$\vert d\rangle$}. These doorways however couple to the
\mbox{$\vert q'\rangle$} states through a strong parity conserving
interaction giving rise to a width of the doorway.

At this stage we introduce an average over states
\mbox{$\vert q'\rangle$}. Since we deal in the present work with
doorways that are of {\em one-particle nature\/} the averaging over
\mbox{$\vert q'\rangle$} states in the effective Hamiltonian
Eq.\ (\ref{6}) will be the same as the averaging which leads to
the appearance of a smoothly varying optical model potential. The
averaging interval $I$ should satisfy \mbox{$D_{sp}>I\gg D_{q}$},
where $D_{sp}$ and $D_{q}$ are the average spacings between the
single-particle states and the compound states
correspondingly. As a result of this averaging the $s$-wave channel
wave functions $\tilde{\Phi}_{s}$ will be solutions of a smooth,
complex optical model potential $\tilde{{\cal H}}_{PP}$. The p-wave
channel wave function is not affected by this average and
$\tilde{\Phi}_{p}\equiv\Phi_{p}$.

Let us now concentrate on Eq.\ (\ref{17}) for the T-matrix and apply
to it the doorway state approximation. Using now $\tilde{\Phi}_{s}$
and taking a Lorentzian average of the part containing the
$\vert q\rangle$ states we arrive at an expression:
\begin{equation}
\tilde{T}_{sp,comp}^{PV}=\sum_{r,q}
\frac{\langle\tilde{\Phi}_{s}^{^{(-)}}\vert H_{PQ}\vert q\rangle
\langle q\vert V^{PV}\vert r\rangle \langle r\vert H_{RP}\vert
\Phi_{p}^{^{(+)}}\rangle}
{(E-E_{r}+\frac{i}{2}\Gamma_{r})(E-H_{QQ}+\frac{i}{2}I)} \ .
\label{37}
\end{equation}
We use now the following relation \cite{Feshbach-Lemmer,Kerman-deTol}:
\begin{eqnarray}
&&D\frac{1}{E-H_{QQ}+\frac{i}{2}I}D= \nonumber \\
&&D\frac{1}{E-H_{DD}+\frac{i}{2}I-
H_{DQ'}[E-H_{Q'Q'}+\frac{i}{2}I]^{-1}H_{Q'D}}D \ .
\nonumber \\
\label{38}
\end{eqnarray}
Assuming random signs for the matrix elements
\mbox{$\langle q\vert H_{Q'D}\vert d\rangle$}
one may write \cite{Feshbach-Lemmer,Auerbach72,Kerman-deTol}:
\begin{equation}
D\frac{1}{E-H_{QQ}+\frac{i}{2}I}D=\sum_{d}\vert d\rangle
\frac{1}{E-E_{d}+\frac{i}{2}\Gamma_{d}^{\downarrow}}\langle d\vert
\label{39}
\end{equation}
with
\begin{equation}
E_{d}-\frac{i}{2}\Gamma_{d}^{\downarrow}=\langle d\vert H\vert
d\rangle +\sum_{q'}\frac{\langle d\vert H\vert q'\rangle
\langle q'\vert H\vert d\rangle}{E-E_{q'}+\frac{i}{2}I}
\label{40}
\end{equation}
so that Eq.\ (\ref{17}) becomes:
\begin{equation}
\tilde{T}_{sp,comp}^{PV}=\sum_{r,d}
\frac{\langle\tilde{\Phi}_{s}^{^{(-)}}\vert H_{PD}\vert d\rangle
\langle d\vert V^{PV}\vert r\rangle \langle r\vert H_{RP}\vert
\Phi_{p}^{^{(+)}}\rangle}{(E-E_{r}+\frac{i}{2}\Gamma_{r})
(E-E_{d}+\frac{i}{2}\Gamma_{d}^{\downarrow})} \ .
\label{41}
\end{equation}
Using now Eq.\ (\ref{3}) and Eq.\ (\ref{41}) we find for the compound
contribution to the asymmetry at resonance energy \mbox{$E=E_{r}$} the
expression:
\widetext
\begin{eqnarray}
P_{comp}(E_{r})=&&-2\sum_{d}\frac{Re(\langle\tilde{\Phi}_{s}^{^{(-)}}
\vert H_{PD}\vert d\rangle\langle d\vert V^{PV}\vert r\rangle\langle r
\vert H_{RP}\vert\Phi_{p}^{^{(+)}}\rangle)
(E_{r}-E_{d})}{\vert\langle r\vert H_{RP}\vert \Phi_{p}^{^{(+)}}\rangle
\vert^{2}[(E_{r}-E_{d})^{2}+\frac{\Gamma_{d}^{\downarrow 2}}{4}]}
\nonumber
\\&&-2\sum_{d}
\frac{
Im(\langle\tilde{\Phi}_{s}^{^{(-)}}\vert H_{PD}\vert d\rangle
\langle d\vert V^{PV}\vert r\rangle \langle r\vert H_{RP}\vert
\Phi_{p}^{^{(+)}}\rangle)
\frac{\Gamma_{d}^{\downarrow}}{2}}
{\vert\langle r\vert H_{RP}\vert \Phi_{p}^{^{(+)}}\rangle\vert^{2}
[(E_{r}-E_{d})^{2}+\frac{\Gamma_{d}^{\downarrow 2}}{4}]} \ .
\label{42}
\end{eqnarray}
\narrowtext
Taking the case when
\mbox{$\vert E_{r}-E_{d}\vert\gg\Gamma_{d}^{\downarrow}$} we
find the expression:
\begin{eqnarray}
&&P_{comp}(E_{r}) \nonumber \\
&&=-2\sum_{d}\frac{Re(\langle\tilde{\Phi}_{s}^{^{(-)}}\vert H_{PD}
\vert d\rangle\langle d\vert V^{PV}\vert r\rangle \langle r\vert
H_{RP}\vert\Phi_{p}^{^{(+)}}\rangle)}
{\vert\langle r\vert H_{RP}\vert \Phi_{p}^{^{(+)}}\rangle\vert^{2}
(E_{r}-E_{d})} \ . \nonumber
\\ \label{43}
\end{eqnarray}
Note that this expression can be obtained from Eq.\ (\ref{25}) by
replacing the states $\vert q\rangle$ with the doorways
$\vert d\rangle$ and the wave function $\Phi_{s}^{^{(-)}}$
with a smooth optical wave function $\tilde{\Phi}_{s}^{^{(-)}}$.
At this stage we have not yet specified the doorways. These will
depend on the nature of the $H$ we use.

One can write Eq.\ (\ref{42}) again in a concise form by making use of
Eq.\ (\ref{32}) and the equation:
\begin{equation}
\gamma_{r}^{(PV)\uparrow}=\sum_{d}\frac
{\langle d\vert V^{PV}\vert r\rangle}{E_{r}-E_{d}+\frac{i}{2}
\Gamma_{d}^{\downarrow}}\gamma_{d}^{\uparrow}(E_{r}) \ ,
\label{44}
\end{equation}
where $\gamma_{d}^{\uparrow}(E_{r})$ is the escape amplitude of the
doorway:
\begin{equation}
\gamma_{d}^{\uparrow}(E_{r})=\langle \tilde{\Phi}_{s}^{^{(-)}}\vert
H_{PD}\vert d\rangle \ .
\label{45}
\end{equation}

Eq.\ (\ref{27}) can be treated in the doorway state approximation by
replacing the states $\vert q\rangle$ by a doorway. However, the
matrix element
\mbox{$\langle d\vert V^{PV}\vert\Phi_{p}(E')\rangle$}
will be much smaller, than the matrix element
\mbox{$\langle d\vert V^{PV}\vert r\rangle$} because the doorway
$\vert d\rangle$ is not related through $V^{PV}$ to
\mbox{$\Phi_{p}(E')$}. We
expect therefore that the contribution in Eq.\ (\ref{27}) will be
a small correction to \mbox{$P_{comp}(E_{r})$} in Eq.\ (\ref{25}).

\section{ONE-BODY PARITY VIOLATING POTENTIAL AND THE SPIN-DIPOLE DOORWAY}
\label{sec:spin-dipole}

In the preceeding sections we have discussed in general terms the
derivation of the longitudinal asymmetry without specifying the nature
of the parity violating part of the Hamiltonian. Therefore it was not
possible to specify the nature of the doorways that might be relevant
in the calculation of the asymmetry or other quantities related to the
experiments. We shall now proceed by limiting our treatment to a
one-body parity violating potential. (In a forthcoming publication we
shall address the question of two-body parity violating interactions
and the question of the relevant doorways in such a case). The
one-body potential we shall now treat is assumed to originate from a
two-body parity violating interaction of the type discussed in
Ref.\ \cite{DDH}. The two-body PV force as discussed in the above
reference or other work is constructed from meson-exchange models and
therefore contains a substantial amount of phenomenological input. The
interactions so constructed do not of course incorporate the very
short range behaviour of the PV interaction. Convoluting such two-body
interactions with the nuclear density one obtains a one-body parity
violating potential. Requiring that in first order such potential will
be even under time reversal one arrives at a potential that has the
form \cite{Michel}:
\begin{equation}
V^{PV}=\epsilon 10^{-7}\frac{1}{2}\sum_{i}
\{ f(r_{i}),\mbox{\boldmath $\sigma_{i}\cdot p_{i}$}c\} \ ,
\label{46}
\end{equation}
where \mbox{\boldmath $\sigma$} is the nucleon spin and
$\mbox{\boldmath $p$}=-i\hbar\mbox{\boldmath $\nabla$}$ the nucleon
momentum operator, $f(r)$ is a radial function and $\epsilon$ is a
parameter that designates the strength of the potential. The curly
brackets denote the anticommutation operation. The weak interaction
does not conserve isospin therefore the parity violating potential may
have an isoscalar and isovector part.

It was demonstrated \cite{AuR92,Kadmensky} that the doorway that
produces a coupling between the state $\vert r\rangle$ and the $\vert
q\rangle$ states is the spin-dipole giant  state \cite{Gaarde} built
on $\vert r\rangle$. This doorway is obtained by acting with the
\begin{equation}
\sum_{i}\mbox{\boldmath $\sigma_{i}\cdot r_{i}$}\sim
\sum_{i} r_{i}[\mbox{\boldmath $\sigma_{i}\otimes
Y_{1}(\hat{r}_{i})$}]^{0^{-}}
\label{47}
\end{equation}
operator on $\vert r\rangle$:
\begin{equation}
\vert SD\rangle_{0}=\frac{1}{N_{0}}\sum_{i} r_{i}
[\mbox{\boldmath $\sigma_{i}\otimes Y_{1}(\hat{r}_{i})$}]^{0^{-}}
\vert r\rangle \ ,
\label{48}
\end{equation}
where $N_{0}$ is a normalization constant, \mbox{\boldmath $Y_{1}$} is
the vector spherical harmonic of order $1$. The spin operator is
treated as a vector and when coupled to the spherical harmonic
\mbox{\boldmath $Y_{1}$} can produce an operator that carries total
spin \mbox{$J=0^{-}$}, $1^{-}$, $2^{-}$. The state in Eq.\ (\ref{48})
is  of isoscalar type and when we multiply the
operator in Eq.\ (\ref{47}) by the isospin operator $\tau_{z}$ and act
on the state $\vert r\rangle$ we obtain the isovector spin-dipole
giant state:
\begin{equation}
\vert SD\rangle_{1}=\frac{1}{N_{1}}\sum_{i} r_{i}
[\mbox{\boldmath $\sigma_{i}\otimes Y_{1}(\hat{r}_{i})$}]^{0^{-}}
\tau_{zi}\vert r\rangle \ ,
\label{49}
\end{equation}
where $N_{1}$ is again a normalization constant.
We should mention here that in fact it is the isovector spin dipole
that was studied quite extensively both experimentally
\cite{Gaarde,Moinester} and theoretically \cite{Auerbach-Klein}. The
information that exists about the isoscalar spin-dipole is rather
scarce \cite{Bland}. In the work \cite{AuR92} it was pointed out that
the spin-dipole giant resonance exhausts most of the strength
associated with the \mbox{\boldmath $\sigma\cdot p$} operator of the
PV potential. (In fact for harmonic oscillator wave functions  this
spin-dipole state exhausts all the strength). The states in
Eqs.\ (\ref{48}) and (\ref{49}) are to be used in the expressions in
Eq.\ (\ref{42}). The dipole giant resonance based on the g.s.\ is
located at the position \mbox{$E\simeq 78A^{-1/3}$ MeV}. Experiments
\cite{Snover} show that the energy position of a giant dipole
resonance built on an excited state is not very different from the one
built on the g.s. Experimental studies indicate
\cite{Gaarde,Moinester} that the spin-dipole is close in energy  to
the spin independent dipole, therefore:
\begin{equation}
E_{d}-E_{r}\simeq 78A^{-1/3} MeV \ .
\label{50}
\end{equation}
The width of the spin-dipole is of the order of a few MeV and thus the
condition \mbox{$\vert E_{d}-E_{r}\vert >\Gamma_{d}^{\downarrow}$} is
valid and Eq.\ (\ref{43}) should be used. We are dealing with the
limit of distant doorways. (In a future publication we shall address
the question of nearby doorways).

We should note the following: in an odd-even nucleus the spin dipole
operator when acting say on a single-particle $p_{1/2}$ component of
the \mbox{$\vert r\rangle$}
state will produce a component which will be a s.p.\ $s_{1/2}$ state.
This component will be only a part of the $\vert SD\rangle$ doorway,
in addition there will be other components which are particle-hole
excitations with respect to state $\vert r\rangle$. (When we say
``single-particle'' we mean a single-particle state with respect to
the target g.s.). In our present approach as well as the
one in Ref.\ \cite{AuR92,Auer-Bow} the doorway is chosen in such
a way as to maximize the matrix element
\mbox{$\langle r\vert V\vert d\rangle$}. We
note however that in the expression in Eq.\ (\ref{43}) appears in the
numerator also the escape amplitude of the doorway. The choice of the
doorway given in Eqs.\ (\ref{48},\ref{49}) does not involve the
consideration to maximize this quantity. In an approach taken in
Ref.\ \cite{Bow-John} the $s_{1/2}$ s.p.\ state is chosen as a
doorway. This choice does maximize $\gamma_{d}^{\uparrow}$ but does
not give consideration to the size of the PV matrix element. Both
approaches give the same result for \mbox{$P_{comp}(E_{r})$}.
Since the final expression was derived in detail in
Ref.\ \cite{Auer-Bow} we shall present here only the resulting
expression for the asymmetry \mbox{$P_{comp}(E_{r})$}. (For reasons of
completeness we repeat the derivation in Appendix \ref{app:2}). The
bound s.p.\ $p_{1/2}$ belonging to the $\{R\}$ space will be denoted
as \mbox{$\vert \phi_{p_{1/2}}^{B}\rangle$}; the bound s.p.\ $s_{1/2}$
belonging to the $\{Q\}$ space will be denoted as \mbox{$\vert
\phi_{s_{1/2}}^{B}\rangle$} and the s.p.\ escape amplitudes will be
denoted as:
\begin{eqnarray}
\gamma_{p}^{\uparrow}(E_{r})=\langle\phi_{p_{1/2}}^{B}\vert H_{RP}\vert
\Phi_{p}^{^{(+)}}\rangle \ ;
\nonumber \\
\gamma_{s}^{\uparrow}(E_{r})=\langle\tilde{\Phi}_{s}^{^{(-)}}
\vert H_{PD}\vert\phi_{s_{1/2}}^{B}\rangle \ .
\label{51}
\end{eqnarray}
The asymmetry is:
\begin{equation}
P_{comp}(E_{r})=-2\frac{Re(\gamma_{s}^{\uparrow}(E_{r})
\langle\phi_{s_{1/2}}^{B}\vert V^{PV}\vert\phi_{p_{1/2}}^{B}\rangle
\gamma_{p}^{\uparrow}(E_{r}))}
{\vert\gamma_{p}^{\uparrow}(E_{r})\vert^{2}\cdot(E_{r}-E_{SD,r})} \ ,
\label{52}
\end{equation}
where $E_{SD,r}$ is an energy, representing the average position of
the isoscalar and isovector spin-dipole, built on the state $\vert
r\rangle$. We should stress that this equation is derived under the
assumption, that $f(r)$ in Eq.\ (\ref{46}) is a constant. The use of a
realistic $f(r)$ will numerically changes the result very little.

The same kind of expression was derived in Ref.\ \cite{Bow-John} where
the doorway as mentioned was the state
$\vert \phi_{s_{1/2}}^{B}\rangle$. The only difference is that in the
denominator of Eq.\ (\ref{52}) $E_{SD,r}$ is the energy position of a
collective spin-dipole excitation while in Ref.\ \cite{Bow-John} the
energy position of the single-particle state
$\vert\phi_{s_{1/2}}^{B}\rangle$ appears. The expression in
Eq.\ (\ref{52}) depends very weakly on the particular structure of the
state $\vert r\rangle$. The only reference to $\vert r\rangle$ is in
the denominator, but \mbox{$E_{r}-E_{SD,r}$} has only a very slight
dependence on $\vert r\rangle$. Of course there is a rather strong
dependence on the energy $E_{r}$ (when $E_{r}$ is small) in the ratio:
\begin{equation}
\frac{\gamma_{s}^{\uparrow}(E_{r})}{\gamma_{p}^{\uparrow}(E_{r})}\sim
\frac{1}{\sqrt{E_{r}}} \ .
\label{53}
\end{equation}
Note, that the sign of Eq.\ (\ref{52}) is fixed for all
$\vert r\rangle$.

\section{PARITY VIOLATING SPREADING WIDTHS}
\label{sec:spread-widths}

In the case of model states, thus states that are not eigenstates of
the full Hamiltonian it is useful to use the notation of a spreading
width. This width is a measure of the extent to which the model state
is not an eigenstate and is spread among the eigenstates of the
system. If a state is a solution of a Schr\"odinger equation with a
model Hamiltonian having some symmetry then one can introduce the
coupling via a symmetry breaking potential of this state to the rest
of space and see how this state is spread out. In this case one can
speak about a symmetry breaking spreading width. In our case of course
it is the reflection symmetry which is relevant. We may ask the
question what is the parity violating spreading width of the negative
parity state in the $\{R\}$ space when coupled to the other two
spaces. In fact the division into spaces is designed so that the
coupling to the $\{Q\}$ space is only through the parity violating
potential. The spreading width for any state $\vert r\rangle$ is given
by \cite{Feshbach,Feshbach-Lemmer,Auerbach72,French,Kerman-deTol}:
\begin{equation}
\Gamma_{r}^{\downarrow}=-2Im\sum_{n}\frac
{\vert\langle r\vert H\vert n\rangle\vert^{2}}{E_{r}-E_{n}+
\frac{i}{2}I} \ ,
\label{54}
\end{equation}
where $I$ is the averaging interval over the $\vert n\rangle$ states,
such that $I\gg D_{n}$, where $D_{n}$ is the average spacing between
states $\vert n\rangle$. The sum extends in principle over all states,
different from $\vert r\rangle$. We shall now consider only the
parity-violating part in $H$ \mbox{$(V^{PV})$} and neglect the
coupling to the $\{P\}$ space. Eq.\ (\ref{54}) becomes:
\begin{equation}
\Gamma_{r}^{\downarrow}=-2Im\sum_{q}\frac
{\vert\langle r\vert H\vert q\rangle\vert^{2}}{E_{r}-E_{q}+
\frac{i}{2}I} \ .
\label{55}
\end{equation}
We may now apply the doorway state approximation to
$\Gamma_{r}^{\downarrow PV}$. Following
Ref.\ \cite{AuR92,Feshbach-Lemmer,Auerbach72} we write:
\begin{equation}
\Gamma_{r}^{\downarrow PV}
=\sum_{d}\frac
{\vert\langle r\vert V^{PV}\vert d\rangle\vert^{2}}
{(E_{r}-E_{d})^{2}+\frac{\Gamma_{d}^{\downarrow 2}}{4}}
\Gamma_{d}^{\downarrow}\ ,
\label{56}
\end{equation}
where the sum is over doorways $\vert d\rangle$ and
$\Gamma_{d}^{\downarrow}$ is the spreading width of the doorway
resulting from the strong interaction coupling of $\vert d\rangle$ to
the $\vert q'\rangle$ states. The expression for
$\Gamma_{d}^{\downarrow}$ is given by
\cite{Feshbach-Lemmer,Auerbach72}:
\begin{equation}
\Gamma_{d}^{\downarrow}=-2Im\sum_{q'}\frac
{\vert\langle d\vert H\vert q'\rangle\vert^{2}}{E_{d}-E_{q'}+
\frac{i}{2}I} \ ,
\label{57}
\end{equation}
In Fig.~\ref{fig:gamma-d-pv} a schematic representation of the doorway
state approach to $\Gamma_{d}^{\downarrow PV}$ is shown. In the case
of the spin-dipole doorways, we have two, the isoscalar and isovector.
Therefore:
\begin{equation}
\Gamma_{r}^{\downarrow PV}
=\frac{\vert\langle r\vert V^{PV}\vert SD_{0}\rangle\vert^{2}}
{(E_{0}-E_{r})^{2}+\frac{\Gamma_{0}^{\downarrow 2}}{4}}
\Gamma_{0}^{\downarrow} +\frac{\vert\langle r\vert V^{PV}\vert SD_{1}
\rangle\vert^{2}}{(E_{1}-E_{r})^{2}+
\frac{\Gamma_{1}^{\downarrow 2}}{4}}\Gamma_{1}^{\downarrow}\ ,
\label{58}
\end{equation}
where $E_{0}$, $E_{1}$, $\Gamma_{0}^{\downarrow}$,
$\Gamma_{1}^{\downarrow}$ are the energy centroids and spreading
widths of isoscalar and isovector spin-dipole giant resonances. In our
numerical estimates we shall replace the two doorways by one which
effectively will represent the two doorways, thus:
\begin{equation}
\Gamma_{r}^{\downarrow PV}
=\frac{\vert\langle r\vert V^{PV}\vert SD\rangle\vert^{2}}
{(\bar{E}_{d}-E_{r})^{2}+\frac{\bar{\Gamma}_{d}^{\downarrow 2}}{4}}
\bar{\Gamma}_{d}^{\downarrow} \ ,
\label{59}
\end{equation}
where $\bar{E}_{d}$, $\bar{\Gamma}_{d}^{\downarrow}$ is the average
energy centroid, and spreading width of the combined distribution of
isoscalar and isovector spin-dipole strength. The spreading width has
the property that is weakly dependent on the density of states,
consequently also on the excitation energy. It can provide therefore a
measure of symmetry breaking. As we shall see in
Sec.~\ref{sec:numcalc} the parity violating spreading width was
determined in the recent experiments \cite{Bowman1,Bowman2} and a
comparison between our estimates and the experimental
$\Gamma_{r}^{\downarrow PV}$ will provide valuable insight into the
possible mechanisms of parity mixing in the compound nucleus.

\section{ORTHOGONALIZATION}
\label{sec:orthog}

When we calculate wave functions belonging to the $\{P\}$ space of
open channels and the wave functions of the $\{Q\}$ or $\{R\}$ spaces,
consisting of bound states we find these by solving the Schr\"odinger
equation with different Hamiltonians. This leads to the loss of
orthogonality between the above spaces. In order to correct for this
one needs to orthogonalize the corresponding wave functions
explicitly. We do this by orthogonalizing the continuum
single-particle wave function $\phi_{p_{1/2}}$ or $\phi_{s_{1/2}}$ to
the corresponding single-particle bound state wave functions found in
the $\{Q\}$ or $\{R\}$ spaces.

The orthogonalization procedure used here is taken from
\cite{Auerbach72}, where it was applied to the isobaric-analog
resonances. Let $h$ be the optical Hamiltonian and $\vert\phi\rangle$
the function that solves the Schr\"odinger equation:
\begin{equation}
h\vert\phi^{^{(+)}}\rangle=E\vert\phi^{^{(+)}}\rangle \ .
\label{60}
\end{equation}
Let us denote by $\vert u\rangle$ the normalized nuclear
single-particle wave function belonging to $\{R\}$ or $\{Q\}$ space.
Then we can write a projected optical Hamiltonian $\tilde{h}$ as:
\begin{equation}
\tilde{h}=(1-\vert u\rangle\langle u\vert)h(1-\vert u\rangle
\langle u\vert) \ .
\label{61}
\end{equation}
The projected wave function $\vert\tilde{\phi}\rangle$ satisfies the
equations:
\begin{equation}
\tilde{h}\vert\tilde{\phi}^{^{(+)}}\rangle=E\vert\tilde{\phi}^{^{(+)}}
\rangle \mbox{ \  and \  }   \langle\tilde{\phi}\vert u\rangle=0 \ .
\label{62}
\end{equation}

To obtain a solution of these equations we use the following method
\cite{Auerbach72}; after solving the homogenous equation (\ref{60}) we
solve the inhomogeneous equation:
\begin{equation}
(E-h)\vert f\rangle=-\vert u\rangle \ ,
\label{63}
\end{equation}
then the function $\vert\tilde{\phi}^{^{(+)}}\rangle$ is given by:
\begin{equation}
\vert\tilde{\phi}^{^{(+)}}\rangle =\lambda\left (\vert\phi^{^{(+)}}
\rangle-\vert f\rangle\frac{\langle u\vert\phi^{^{(+)}}\rangle}
{\langle u\vert f\rangle}\right ) \ .
\label{64}
\end{equation}
Here the constant $\lambda$ is determined by requirement of the usual
asymptotic normalization of $\vert\tilde{\phi}^{^{(+)}}\rangle$.

In Sec.~\ref{sec:numcalc} we shall discuss the numerical results and
the effect of the orthogonalization on the results.

\section{CALCULATIONS AND ESTIMATES}
\label{sec:numcalc}

In this section we shall evaluate numerically (or make order of
magnitude estimates in the case of small terms) the various
contributions to the asymmetry \mbox{$P(E_{r})$} derived in
Sec.~\ref{sec:asymm} as well as the PV spreading width discussed in
Sec.~\ref{sec:spread-widths}.

Before we deal with each of the terms separately we present the
general ingredients used in our calculations. We shall perform all the
calculations for the compound nucleus of $^{233}Th$, considering it as
a typical case. Most of the results that will be calculated for this
nucleus should apply also to other nuclei in this region of the
periodic table, such as  U  etc. There is however one peculiarity to
$^{233}Th$, namely that all the nine experimental asymmetries that are
statistically significant turn out to be positive in this nucleus
\cite{Bowman2}. This kind of sign correlation was not found in the
other nuclei for which experiments were performed. We shall return to
this point later.

\subsection{PV interaction and optical potential}
\label{subsec:optpot}

For the effective one-body PV interaction we used the potential in
Eq.\ (\ref{46}). The radial function $f(r)$ has a Woods-Saxon shape
with the same parameters as $f_{a}(r)$ used in the optical potential
Eq.\ (\ref{65}) below.

The continuum wave functions were calculated using the Perey-Buck
\cite{Perey} and Madland-Young \cite{Hussein,Hod85} optical
potentials, and the computer code DWUCK4 \cite{Kunz}. Both potentials
have the form:
\begin{equation}
V_{opt}=V_{0}f_{a}(r)+V_{ls}r_{ls}^{2}\frac{1}{r}\frac{df_{ls}}{dr}
\mbox{\boldmath $ls$}+
iW_{0}a_{b}\frac{df_{b}}{dr} \ ,
\label{65}
\end{equation}
where \mbox{$f_{i}(r)\equiv[1+exp((r-R)/a_{i})]^{-1}$} is the
Woods-Saxon function, \mbox{$R=r_{0}A^{1/3}$}. The parameters of the
Perey-Buck potential are:
\mbox{$V_{0}=-48.0$} MeV, \mbox{$V_{ls}=14.8$} MeV, \mbox{$W_{0}=44$}
MeV, \mbox{$r_{0}=r_{ls}=1.27$} fm, \mbox{$a_{a}=a_{ls}=0.65$} fm,
\mbox{$a_{b}=0.47$} fm. The parameters of the Madland-Young potential
are: \mbox{$V_{0}=-(50.378-27.073\frac{N-Z}{A}-0.345E_{lab})$} MeV,
\mbox{$V_{ls}=24.3$} MeV,
\mbox{$W_{0}a_{b}=4(9.265-12.666\frac{N-Z}{A}-0.232E_{lab})$}
\mbox{MeV$\cdot$fm }, \mbox{$r_{0}=1.264$} fm, \mbox{$r_{ls}=1.01$}
fm, \mbox{$a_{a}=0.612$} fm, \mbox{$a_{b}=0.553+0.0144E_{lab}$} fm,
\mbox{$a_{ls}=0.75$} fm. Here $E_{lab}$ is the neutron energy in MeV.
The depth parameters of volume, spin-orbit and imaginary surface
potentials  for $^{232}Th$ and \mbox{$E_{lab}=1$ eV} are:
the Perey-Buck potential, \mbox{$V_{0}=-48.0$} MeV,
\mbox{$V_{ls}r_{ls}^{2}=23.9$} \mbox{MeV$\cdot$fm$^{2}$},
\mbox{$W_{0}b=20.7$} \mbox{ MeV$\cdot$fm};
the Madland-Young potential, \mbox{$V_{0}=-44.3$} MeV,
\mbox{$V_{ls}r_{ls}^{2}=24.8$} \mbox{MeV$\cdot$fm$^{2}$},
\mbox{$W_{0}b=14.2$} \mbox{MeV$\cdot$fm}.

In our formalism the optical model operator
(Sec.~\ref{sec:react-theory}) has a different structure in the case of
the $s$ and $p$ channels, which is reflected in fact, that we do not
average over the $p_{1/2}$, but average over the $s_{1/2}$ resonances.
Because of that in our calculations of the $p_{1/2}$ continuum wave
functions we do not include the imaginary part of the optical
potential. In our studies we found, however, that while the asymmetry
$P(E)$ is sensitive to the depth of the real part of the optical
potential, it is not sensitive to the presence or absence of the
imaginary part of the potential in either the $s$ or $p$-wave
channels.

The relevant single-particle bound states in the $Th$ region are the
$4p_{1/2}$,  $4s_{1/2}$, $5s_{1/2}$. In Sec.~\ref{sec:spin-dipole}
they were denoted as \mbox{$\vert \phi_{p_{1/2}}^{B}\rangle$} and
\mbox{$\vert \phi_{s_{1/2}}^{B}\rangle$}. The wave functions of these
states were calculated in a Woods-Saxon potential:
\begin{equation}
V_{sm}=V_{1}f_{s}(r)+V_{1}\xi r_{0}^{2}\frac{1}{r}\frac{df_{s}}{dr}
\mbox{\boldmath $ls$} \ ,
\label{66}
\end{equation}
where $a_{s}=0.60$ fm, $r_{s}=1.29$ fm, \mbox{$\xi=-0.804$} and
\mbox{$V_{1}=-69.7$ MeV}. This set is standard in calculations for
nonspherical nuclei such as $Th$ \cite{Chasman}. This potential
produces a deeply bound $5s_{1/2}$ state, which in nature is in fact
unbound. The reason is that when this potential is used in deformed
nuclei, additional terms are added. These terms are not included in
the present calculations. To make the state $5s_{1/2}$ weakly bound we
also used a second set with \mbox{$V_{1}=-59.5$ MeV}.

In our calculations we checked the sensitivity of the asymmetry $P(E)$
to the depth parameters of the optical and shell model potentials to
verify the stability of the numerical results.

The strong interaction parts of the Hamiltonians $H_{PR}$ and $H_{QP}$
contain both one and two-body parts. However as a consequence of the
fact, that in the present approach we study only the one-body PV
potential the contributions of the two-body parts of $H_{PR}$ and
$H_{QP}$ mostly drop out. In some cases the two-body will contribute.
This is however beyond the scope of the present work and it will be
discussed in forthcoming work.

We use for these two interactions a real Woods-Saxon potential
and denote it $V_{str}$. For convenience the parameters of these
transition potentials are taken from the real part of the
of the Perey-Buck potential. We should remark that these are not
optical potentials which are used in the calculation of the channels
that are in the $\{P\}$ space. We should also emphasize that in our
approach the one-body potentials are not represented by
self-consistent Hartree-Fock potentials because for example the
$\{R\}$ space contains only bound single-particle components while the
unbound components are in the $\{P\}$ space which is orthogonal to
$\{R\}$. This is inherent in the nature of the projection operator
approach \cite{Feshbach,Feshbach-Lemmer,Auerbach72}.

\subsection{Effect of orthogonalization}
\label{subsec:orthog-eff}

When we orthogonalize the continuum wave function (w.f.) to the
single-particle bound wave function, we can see, that the number of
nodes and, hence, the phase at the origin of the orthogonalized
function is determined by the behaviour of the bound wave function.
(Of course the asymptotic behaviour of the orthogonalized wave
function should not be changed). The behaviour of orthogonalized and
non-orthogonalized continuum radial wave functions is illustrated in
Fig.~\ref{fig:rad-ort-no-wf}. The continuum $s$ and $p$ w.f., when
orthogonalized to the bound $4p_{1/2}$ and $5s_{1/2}$ radial wave
functions change the phase at the origin, because they acquire an
additional node. The $s$-wave continuum w.f., orthogonalized to the
$4s_{1/2}$ radial w.f.\ is just distorted so, that the spacing between
nodes is compressed. This explains the changing of the phases of the
single-particle amplitudes. The strong effect of orthogonalization in
the present calculations is due to the fact that the continuum wave
functions are very close to threshold. (This is quite different from
the case of analog resonances \cite{Auerbach72}).

Because we are considering here neutrons with energies
\mbox{$E_{n}\sim 1-500$ eV}, which are much smaller than the depth of
the nuclear potential, the behaviour of the continuum wave functions
inside a nucleus is determined by the potential and is weakly
dependent on the neutron energy. All the terms, which contribute to
$P(E)$ have the same energy dependence \mbox{$1/\sqrt{E}$} due to the
factor \mbox{$1/\gamma_{p}$}.

We shall consider as representative the calculation with
orthogonalized continuum wave functions in the Perey-Buck optical
potential and with bound states calculated in the potential with
$V_{1}=-69.7$ MeV.

\subsection{Calculation of $P_{comp}(E_{r})$}
\label{subsec:P-comp}

We calculate $P_{comp}(E_{r})$ in the doorway state approximation. The
formula in Eq.\ (\ref{52}) for the on-resonance asymmetry
\mbox{$P_{comp}(E_{r})$} involves the matrix element of $V^{PV}$
between s.p.\ bound states and two s.p.\ escape amplitudes. Our
calculation gives for the matrix elements
\mbox{$\langle 4p_{1/2}\vert V^{PV}\vert ns_{1/2}\rangle$}
(in units of $\epsilon$) \mbox{$-i17.2$ eV} and \mbox{$-i17.0$ eV} for
the deeply bound $4s$ and $5s$ states respectively and $-i15.8$,
\mbox{$-i10.3$ eV} for the more weakly bound. We used here the phase
convention of positive radial wave functions near the origin. The
corresponding matrix elements for oscillator wave functions are
\mbox{$-i16.1$ eV} and \mbox{$-i15.2$ eV}.
Using Eq.\ (\ref{52}) we obtain:
\widetext
\begin{eqnarray}
P_{comp}(E_{r})=&&-2\frac{Re(\langle\tilde{\Phi}_{s}^{^{(-)}}
\vert V_{str}\vert 4s_{1/2}\rangle\langle 4s_{1/2}\vert V^{PV}
\vert 4p_{1/2}\rangle\langle 4p_{1/2}\vert V_{str}\vert
\Phi_{p}^{^{(+)}}\rangle )}{\vert\langle 4p_{1/2}\vert
V_{str}\vert \Phi_{p}^{^{(+)}}\rangle\vert^{2}\cdot (E_{r}-E_{d-})}
\nonumber \\
&&-2\frac{Re(\langle\tilde{\Phi}_{s}^{^{(-)}}
\vert V_{str}\vert 5s_{1/2}\rangle\langle 5s_{1/2}\vert V^{PV}
\vert 4p_{1/2}\rangle\langle 4p_{1/2}\vert V_{str}\vert
\Phi_{p}^{^{(+)}}\rangle )}
{\vert\langle 4p_{1/2}\vert V_{str}\vert \Phi_{p}^{^{(+)}}\rangle
\vert^{2}\cdot (E_{r}-E_{d+})} \ .
\label{67}
\end{eqnarray}
\narrowtext
This equation represents the fact, that two doorway states with
$5s_{1/2}$ and $4s_{1/2}$ single-particle w.f.\ can contribute. The
denominators of these two parts have different signs. In our
calculation we used for the energy difference in the denominator the
values of \mbox{$\pm 5$ MeV}, which is supposed to represent the
centroid of the isovector and isoscalar spin-dipole doorways. The
resulting $P_{comp}(E_{r})$ as a function of $E_{r}$ is shown in
Fig.~\ref{fig:doorw-P(E)}.

For the orthogonalized continuum waves both terms in Eq.\ (\ref{67})
have negative sign and the quantity
\mbox{$P_{comp}(E_{r}=1eV)/\epsilon=-0.95\%$}. We found in our
computations that the presence or absence of the imaginary part of the
optical potential in the $s$-wave or in the $p$-wave channel does not
change the results noticeably. We also checked the sensitivity of
\mbox{$P_{comp}(E_{r})$} to the optical and shell model potentials and
the orthogonalization of continuum wave functions. The asymmetry,
calculated with the orthogonalized continuum w.f.\ is sensitive to the
details of bound wave functions, thus to the depth of the shell model
potential.  When $V_{1}$ is changing from $-69.7$ to $-59.5$~MeV the
asymmetry becomes about $1.5$ times smaller for $4s$ term and about
$4$ times smaller for $5s$ term and the final result is:
\mbox{$P_{comp}(E_{r}=1eV)/\epsilon =-0.44\%$}. The sensitivity to the
depth of the optical potential is less, than $10\%$. The ratio
\mbox{$\displaystyle\vert\frac{\gamma_{s}}{\gamma_{p}}\vert$} of
single-particle escape amplitudes varies between
\mbox{$(0.35-0.73)\times10^{3}$}, depending on the number of nodes and
shell model parameters. The square well estimate is
\mbox{$\displaystyle\vert\frac{\gamma_{s}}{\gamma_{p}}\vert\simeq
\frac{\sqrt{3}}{kR}=1\times10^{3}$} \cite{Bohr-Mottelson}.

For the non-orthogonalized continuum waves the two terms in
Eq.\ (\ref{67}) have different signs as opposed to the case of
orthogonalized waves. The asymmetry, calculated with
non-orthogonalized waves, is very sensitive to the depth of real part
of the optical potential. The quantity
\mbox{$P_{comp}(E_{r}=1eV)/\epsilon =0.06\%$}, when calculated with
the Perey-Buck potential \mbox{($V_{0}=-48.0$ MeV)} and $0.34\%$, when
the Madland-Young potential \mbox{($V_{0}=-44.3$ MeV)} was used. The
sensitivity to the depth parameter of the shell-model potential is
smaller, about $20\%$. The ratio
\mbox{$\displaystyle\vert\frac{\gamma_{s}}{\gamma_{p}}\vert=
0.06-0.67\times10^{3}$}.

\subsection{Calculation of $P_{dir}(E_{r})$}
\label{subsec:P-dir}

Assuming, that only the one-body part of $V^{PV}$ or $H_{PR}$ is
contributing and writing the wave function
\begin{equation}
\vert r\rangle=a_{r}\vert 4p_{1/2}\rangle+...
\label{68}
\end{equation}
we obtain from Eq.\ (\ref{24}) that
\begin{equation}
P_{dir}(E_{r})=-2\frac{Re(\langle\tilde{\Phi}_{s}^{^{(-)}}\vert V^{PV}
\vert 4p_{1/2}\rangle\langle 4p_{1/2}\vert V_{str}\vert
\Phi_{p}^{^{(+)}}\rangle)}{\vert\langle 4p_{1/2}\vert
V_{str}\vert \Phi_{p}^{^{(+)}}\rangle\vert^{2}}\ .
\label{69}
\end{equation}
The resulting asymmetry $P_{dir}(E_{r})$ as a function of the neutron
energy is shown in Fig.~\ref{fig:direct-P(E)}. For
\mbox{$E_{r}=1$ eV}, $P_{dir}(E_{r})/\epsilon=0.16\%$. We see, that
the direct term is smaller than the compound term by about a factor of
$6$. The dependence on the depth parameters and orthogonalization is
of the same type, as found in the case of the compound (doorway) term.
When \mbox{$V_{1}=-59.5$ MeV},
\mbox{$P_{dir}(E_{r}=1eV)/\epsilon=0.12\%$}. For the
non-orthogonalized continuum wave function the asymmetry has a
negative sign and \mbox{$P_{dir}(E_{r})/\epsilon=-0.015\%$} in case of
the Perey-Buck potential and $-0.05\%$ for the Madland-Young
potential.

\subsection{Channel coupling resonance terms}
\label{subsec:P-chan-coupl}

The term in Eq.\ (\ref{26}), as was shown in \cite{Lew-Weid}, contains
the enhancement, caused by off-shell contributions
\mbox{$E'\gg E_{r}$}, because at small energies
\mbox{$\phi_{p}(E,r)\sim\sqrt{k\mu}(kr)$} and the integrand is small
for \mbox{$E'=E_{r}$}. The energy dependence of the asymmetry
\mbox{$P_{cc}(E_{r})$} is also \mbox{$1/\sqrt{E_{r}}$}. For
\mbox{$E_{r}=1$ eV}, \mbox{$P_{cc}(E_{r})/\epsilon=-0.15\%$}, while
the unenhanced on-shell result is \mbox{$1\times 10^{-10}\%$}. This
asymmetry is not very sensitive to the parameters of the optical
potential but shows sensitivity to the depth of the shell model
potential:
\mbox{$P_{cc}(E_{r})/\epsilon=-0.32\%$}, when
\mbox{$V_{1}=-59.5$ MeV}. For the non-orthogonalized continuum w.f.\
the asymmetry \mbox{$P(E_{r})/\epsilon=-0.094\%$} in case of the
Perey-Buck optical potential and $-0.07\%$, when the Madland-Young
potential with a constant \mbox{$V_{0}=-44.3$ MeV} was used.

It was suggested in \cite{Lew-Weid} to use a surface-peaked
approximation, when dealing with integrals of the kind, found in
Eq.\ (\ref{26}). The assumption is, that the strong and
parity-violating interactions are surface-peaked and that the
equations
\begin{equation}
\frac{\langle\Phi_{p}(E')\vert H_{PR}\vert r\rangle}{\phi_{p}(k'R)}=
\frac{\langle\Phi_{p}(E)\vert H_{RP}\vert r\rangle}{\phi_{p}(kR)}
\label{70}
\end{equation}
and
\begin{equation}
\frac{\langle\tilde{\Phi}_{s}^{^{(-)}}(E_{r})\vert V^{PV}\vert
\Phi_{p}(E')\rangle}{\phi_{p}(k'R)}=\frac{\langle
\tilde{\Phi}_{s}^{^{(-)}}(E_{r})\vert V^{PV}
\vert\Phi_{p}(E)\rangle}{\phi_{p}(kR)}
\label{71}
\end{equation}
hold through the entire virtual energy scale. With the approximation
\mbox{$E_{r}\cong 0$} (Ref.~\cite{Lew-Weid}):
\begin{eqnarray}
P_{cc}(E_{r})=&&2Re\bigl(\langle\tilde{\Phi}_{s}^{^{(-)}}(E_{r})\vert
V^{PV}\vert\Phi_{p}(E_{r})\rangle\bigr) \nonumber \\ \times &&
\int_{0}^{\infty}\frac{dE'}{E'}\frac{{\vert\phi_{p}(E')\vert}^2}
{{\vert\phi_{p}(E_{r})\vert}^2} \ .
\label{72}
\end{eqnarray}
The enhancement here arises due to the off-shell, effect, i.e.\ when
\mbox{$E'\gg 0$}.

In our calculations we found, that the main contribution to the
integral comes from the interval \mbox{$E'=0.1-100$ MeV}
(see \mbox{Fig.~\ref{fig:surf-peak-int}(a))} both in the exact
calculation and also when the above approximation is applied. However,
because of the rather complex dependence of the continuum wave
function \mbox{$\phi_{p}(E',r)$} on energy, the approximations of
Eq.\ (\ref{70},\ref{71}) are good only in the energy interval
\mbox{$0-1$ MeV} in the case of orthogonalized and
\mbox{$0-0.015$ MeV} in the case of the non-orthogonalized continuum
wave functions. In \mbox{Fig. \ref{fig:surf-peak-int}(b)} we show the
ratio of the product of the left hand side of Eqs.\ (\ref{70}) and
(\ref{71}) of the product of the right hand side of these equations
for the case of \mbox{$E=1$ eV} as a function of the virtual energy
$E'$. We can see that for $E'$ smaller than $1$~MeV the ratio is
approximately one, and thus the surface-peaked approximation works
well. For higher $E'$ energies the approximation breaks down. Note
that the very sharp structure is due to the use of a logarithmic
energy scale. This conclusion does not change when the derivative of a
Woods-Saxon function \mbox{$-f'_{a}(r)$}, which is surface-peaked, was
used for the strong interaction $H_{RP}$.

The value of the asymmetry in the case of the orthogonalized continuum
wave functions and surface-peaked approximation for \mbox{$E_{r}=1$ eV}
is \mbox{$P_{cc}(E_{r})/\epsilon=-0.17\%$}, which is close to the
value obtained in the exact calculation. The agreement might be
fortuitous due to the oscillations of the exact integrand. This result
is not very sensitive to the depth parameter of the optical potential.
The calculation in the surface-peaked approximation with
non-orthogonalized continuum wave functions gives \mbox{$0.61\%$}. As
opposed to the previous orthogonalized continuum case, the latter
result is very sensitive to the depth of the optical potential if
\mbox{$V_{0}>46$ MeV}. In Ref.~\cite{Lew-Weid} the value of
\mbox{$P_{cc}(E_{r}=1eV)/\epsilon =0.25-0.75\%$} is found for this
asymmetry term.

The second channel coupling resonance term in Eq.\ (\ref{27}) will be
small because of the argument given at the end of
Sec.~\ref{sec:doorway}.

\subsection{Channel coupling background terms}
\label{subsec:chan-coupl-backgr}

As pointed out in Sec.~\ref{sec:asymm} the term in Eq.\ (\ref{29}) is
equal to zero in the limit of real radial continuum wave functions
$\phi_{s}$ and $\phi_{p}$. Because
\mbox{$\displaystyle\frac{Im\phi_{s}}{Re\phi_{s}}\sim5\times10^{-2}$}
and
\mbox{$\displaystyle\frac{Im\phi_{p}}{Re\phi_{p}}\sim5\times10^{-9}$}
this term is much smaller, than the one in Eq.\ (\ref{26}). The same
is true for the term in Eq.\ (\ref{31}), which is much smaller, than
the term in Eq.\ (\ref{27}). \\

We summarize all the results in Table \ref{table1}. The values
\mbox{$P(E_{r}=1eV)/\epsilon$}, calculated for different terms are
shown for the case of orthogonalized continuum wave functions and for
the two shell-model potentials used in our calculation of the bound
states.

\subsection{Estimate of the parity violating spreading width}
\label{subsec:est-spreadwidth}

We use Eq.\ (\ref{59}) to estimate the parity violating spreading
width in $^{233}Th$. The matrix element \mbox{$\langle r\vert
V^{PV}\vert d\rangle$} where \mbox{$\vert d\rangle$} is the collective
spin-dipole when compared to a single-particle matrix elements it is
enhanced by a factor $\sqrt{N}$, where $N$ is roughly the number of
particle-hole states, composing the spin-dipole. In the case of
\mbox{$A\simeq 230$}, counting the number of particle-hole states,
\mbox{$N\simeq 100$} and we therefore can use the estimate
\mbox{$\langle r\vert V^{PV}\vert SD\rangle\simeq 10\times
\langle\phi_{p}^{B}\vert V^{PV}\vert\phi_{s}^{B}\rangle$}. For
\mbox{$\epsilon =1$} this gives a matrix element
\mbox{$\langle r\vert V^{PV}\vert SD\rangle\simeq 150$ eV}. Let us now
take \mbox{$E_{d}-E_{r}\simeq 5$ MeV} and
\mbox{$\bar{\Gamma}_{d}^{\downarrow}\simeq 3$ MeV} assuming that this
number represents a spreading width common to the isoscalar and
isovector spin-dipole. A typical spreading width of a giant resonance
is of this size. As for the energy difference the dipole state is at
an excitation energy of about 10~MeV, however we expect the isoscalar
spin-dipole to be lower and therefore we chose the above value for
\mbox{$E_{d}-E_{r}$}. We should emphasize however that these numbers
are used only in an order of magnitude estimate that we make here.
We obtain for
\mbox{$\Gamma_{r}^{\downarrow PV}\simeq 2\times 10^{-3}$ eV}, three
orders of magnitude larger than the experimental value \cite{Bowman2}
of \mbox{$\Gamma_{r}^{\downarrow PV}\simeq 6\times 10^{-7}$ eV}. In
order to obtain agreement one must reduce $\epsilon$ to a value of
about $0.02$. Although the above estimate is rough it clearly
indicates that there is an internal inconsistency in the evaluation of
the PV spreading width and average asymmetry. The same PV matrix
element cannot reproduce simultaneously the average asymmetry and the
spreading width. The later requires a parity violating potential that
has a value of $\epsilon$ which is ``reasonable'' compared to the
values derived using meson-exchange models \cite{DDH}.

\section{DISCUSSION OF RESULTS AND CONCLUSIONS}
\label{sec:discussion}

The calculations of the average asymmetry $P(E_{r})$ in the framework
of the one-body approximation and for realistic \mbox{$\epsilon
=0.1-0.3$} give numbers which are of the order of \mbox{$0.1\%$} or
less. We have considered several contributions, some new, as for
example \mbox{$P_{dir}(E_{r})$}. We have evaluated the channel
coupling term, suggested in Ref.~\cite{Lew-Weid}, exactly, avoiding
the surface-peaked approximation. This term turned out to be of the
order of \mbox{$0.02\%$}, if the above value of $\epsilon$ is used.
Even if all these contributions add coherently the asymmetry found,
would not exceed \mbox{$0.2\%$}. The average asymmetry found in the
experiments for $Th$ and other nuclei is of the order of several
percent, thus at least one order of magnitude smaller than the present
theoretical results. Moreover, when using our doorway model to
estimate the spreading width we find that the matrix element
required to reproduce the experimental PV spreading width is of the
order of a fraction of an eV which corresponds to a value of
\mbox{$\epsilon =0.02$}. Saying it in other words, if we use the value
of $\epsilon$ that fits the asymmetry \mbox{$P(E_{r})$} we shall
overestimate the spreading width by about two orders of magnitude. We
feel that we have exhausted most of the contributions to
\mbox{$P(E_{r})$} in the framework of one-body theories. The above
difficulty to reproduce the correct order of magnitude of the average
asymmetry with reasonable values for $\epsilon$ and the internal
contradiction within the model between the estimate of the spreading
width and the calculation of \mbox{$P(E_{r})$} leads us to the
conclusion that the one-particle model is not sufficient in trying to
account for all the aspects of parity violation effects in the
compound nucleus. One should attempt to go beyond the one-particle
approximation and try to introduce the effects of the residual
two-body force. In that case questions of intermediate structure
\cite{Feshbach-Lemmer,Mekjian} in the nucleus become relevant. In a
future publication we shall address this question.

It was emphasized in the literature
\cite{Bowman2,Bow-John,Lew-Weid,Auer-Bow} that the measured
asymmetries for the individual resonances exhibit sign correlations in
$^{233}Th$. When averaged over all the observed resonances
$\vert r\rangle$ the averaged \mbox{$P(E_{r})$} turns out to be around
\mbox{$8\%$}. We have seen that most of the terms we considered here,
the doorway state approximation to the compound term, the direct term,
as well as the channel coupling term, have signs independent on  the
particular compound resonance $\vert r\rangle$. The problem of course
is that the size of these terms for the realistic values of the PV
coupling constant is, as said above, at least an order of magnitude
smaller than the value found in the $^{233}Th$ experiment
\cite{Bowman2}.

In summary, in this work using a unified reaction theory approach we
have derived in a systematic way all the terms that contribute to the
non-fluctuating part of the longitudinal asymmetry \mbox{$P(E)$}.
Off-resonance and on-resonance terms were identified. All these terms
were treated numerically either by direct calculation or in the case
of small terms, using estimates. The projection operator reaction
theory we used is flexible enough so that by choosing appropriately
the subspaces of the full space we were able to reduce the number of
terms in the description of the asymmetry. The compound term emerged
as the most important one. This term was calculated using the doorway
state approximation with the doorway being the giant spin-dipole
resonance.

\acknowledgments

We thank David Bowman for very helpful discussions and P.D. Kunz for
making avaliable to us the DWUCK code which was used extensively in
the numerical calculations. This work was supported by the US-Israel
Binational Science Foundation.

\appendix

\section{\protect\\ $T$-MATRIX AND ASYMMETRY $P(E)$}
\label{app:1}

The asymmetry $P(E)$ is defined as:
\begin{equation}
P(E)=\frac{\sigma^{^{+}}_{p_{1/2}}(E)-\sigma^{^{-}}_{p_{1/2}}(E)}
{2\sigma^{0}_{p_{1/2}}(E)} \ ,
\label{a1}
\end{equation}
where $+$ and $-$ denote helicities $\nu$ of projectile neutrons. In
this Appendix we derive the relation between the $T$-matrix and
\mbox{$P(E)$}, given in Eq.\ (\ref{3}).
Writing the optical theorem for helicity $\nu$\ :
\begin{equation}
\mbox{$\frac{4\pi}{k}$}
Imf_{\nu\nu}(\vec{k}\rightarrow\vec{k})=\mbox{$\sum_{\nu'}$}
\sigma_{\nu'\nu}
\label{a2}
\end{equation}
we get:
\begin{equation}
\sigma^{^{+}}-\sigma^{^{-}}=\mbox{$\frac{4\pi}{k}$}
(Imf_{++}(0)-Imf_{--}(0)) \ .
\label{a3}
\end{equation}
The partial wave expansion for helicity amplitudes
\mbox{$f_{\nu'\nu}(\theta)$} is \cite{Newton}:
\begin{equation}
f_{\nu'\nu}(\theta)=\mbox{$\sum_{j}$}(2j+1)d^{j}_{\nu'\nu}(\theta)
f^{j}_{\nu'\nu} \ ,
\label{a4}
\end{equation}
where $f^{j}_{\nu'\nu}$ are partial amplitudes. In the
case when \mbox{$j=\frac{1}{2}$} only and \mbox{$\theta=0$}:\
\mbox{$f_{++}(0)=2f_{++}^{j=1/2}$},\
\mbox{$f_{--}(0)=2f_{--}^{j=1/2}$} and
\begin{equation}
\sigma^{^{+}}-\sigma^{^{-}}=\mbox{$\frac{8\pi}{k}$}
(Imf_{++}^{j=1/2}-Imf_{--}^{j=1/2})\ .
\label{a5}
\end{equation}
We go now to the $lsj$ representation, assuming rotational invariance:
\begin{equation}
\vert Ejm_{j}\nu\rangle =\sum_{l}\mbox{$\sqrt{\frac{2l+1}{2j+1}}$}
\langle l0\mbox{$\frac{1}{2}$}\nu
\vert j\nu\rangle \vert Els=\mbox{$\frac{1}{2}$};jm_{j}\rangle \ .
\label{a6}
\end{equation}
The expression for helicity partial amplitudes in terms of partial
amplitudes in the \mbox{$lsj$} representation
\mbox{$f_{l'\frac{1}{2},l\frac{1}{2}}^{j}\equiv f_{l',l}^{j}$} is:
\begin{equation}
f^{j}_{\nu'\nu}=\sum_{l'l}\mbox{$\frac{\sqrt{(2l+1)(2l'+1)}}{2j+1}$}
\langle l'0\mbox{$\frac{1}{2}$}\nu'\vert j\nu'\rangle
\langle l0\mbox{$\frac{1}{2}$}\nu\vert j\nu\rangle
f_{l'\frac{1}{2},l\frac{1}{2}}^{j} \ .
\label{a7}
\end{equation}
Here \ \mbox{$l,l'=j\pm\frac{1}{2}$}\/ and
\begin{eqnarray}
f_{++}^{j}=\mbox{$\frac{1}{2}$}(f_{j-\frac{1}{2},j-\frac{1}{2}}^{j}+
f_{j+\frac{1}{2},j+\frac{1}{2}}^{j}-2f_{j+\frac{1}{2},j-
\frac{1}{2}}^{j})\ ,
\label{a8} \\
f_{--}^{j}=\mbox{$\frac{1}{2}$}(f_{j-\frac{1}{2},j-\frac{1}{2}}^{j}+
f_{j+\frac{1}{2},j+\frac{1}{2}}^{j}+2f_{j+\frac{1}{2},j-
\frac{1}{2}}^{j}) \ .
\nonumber\end{eqnarray}
Because of time-reversal invariance
\mbox{$f_{l',l}^{j}=f_{l,l'}^{j}$}. Then for the case of
\mbox{$j=\frac{1}{2}$}\ :
\begin{eqnarray}
\sigma^{^{+}}-\sigma^{^{-}}=&&\mbox{$\frac{8\pi}{k}$}
(Imf_{++}^{j=1/2}-Imf_{--}^{j=1/2}) \nonumber \\
=&&-\mbox{$\frac{16\pi}{k}$}Imf_{l'=1,l=0}^{j=\frac{1}{2}} \ .
\label{a9}
\end{eqnarray}
Expressing the $T$-matrix elements in terms of partial amplitudes
\mbox{$T_{l'l}^{j}=-\frac{k}{\pi}f_{l',l}^{j}$} we obtain:
\begin{equation}
P(E)=\frac{\sigma^{^{+}}_{p_{1/2}}(E)-\sigma^{^{-}}_{p_{1/2}}(E)}
{\sigma^{0}_{p_{1/2}}(E)}=
\frac{8\pi^{2}}{k^{2}}\frac{ImT_{l'=1,l=0}^{j=\frac{1}{2}PV}(E)}
{\sigma^{0}_{p_{1/2}}(E)} \ .
\label{a10}
\end{equation}

\section{\protect\\ DERIVATION OF EQUATION (\ref{52}) FOR
$P(E_{\lowercase{r}})$}
\label{app:2}

In Eqs.\ (\ref{47},\ref{48},\ref{49}) we specified the isoscalar and
isovector doorways \mbox{$\vert d\rangle$} as:
\begin{eqnarray}
\vert SD\rangle_{0}=\frac{1}{N_{0}}\sum_{i=1}^{A}
(\mbox{\boldmath $\sigma r$})_{i}\vert r\rangle \ ,
\label{b1} \\
\vert SD\rangle_{1}=\frac{1}{N_{1}}\sum_{i=1}^{A}
(\mbox{\boldmath $\sigma r$})_{i}\tau_{zi}\vert r\rangle \ .
\nonumber\end{eqnarray}
In what follows we shall consider one doorway state \mbox{$\vert
d\rangle$} for each \mbox{$\vert r\rangle$} with the normalization
constant $N$. For the sake of simplicity we shall also take the
one-body parity-violating potential in Eq.\ (\ref{46}) in its
schematic form
\mbox{$\displaystyle V^{PV}=g\sum_{i=1}^{A}
(\mbox{\boldmath $\sigma p$})_{i}$}
with $g$ being a constant. The product of matrix elements of $V^{PV}$
and $H_{DP}$ operators in the numerator of Eqs.\ (\ref{41},\ref{43})
is now:
\begin{eqnarray}
&&\langle\tilde{\Phi}_{s}^{^{(-)}}\vert H_{PD}\vert d\rangle
\langle d\vert V^{PV}\vert r\rangle \nonumber \\
&&= \frac{g}{N^{2}}
\langle\tilde{\Phi}_{s}^{^{(-)}}\vert H_{PD}\vert d\rangle
\langle r\vert
\sum_{i=1}^{A}(\mbox{\boldmath $\sigma r$})_{i}
\sum_{i=1}^{A}(\mbox{\boldmath $\sigma p$})_{i}\vert r\rangle \ .
\label{b2}
\end{eqnarray}
The matrix elements of the operator \mbox{\boldmath $\sigma p$} can be
approximately replaced (for example, using the basis of oscillator
wave functions) by the matrix elements of
\mbox{$i\omega m\mbox{\boldmath $\sigma r$}$} (see for example
Ref.\ \cite{Bohr-Mottelson}). (In a recent microscopic calculation
\cite{Auerbach-Brown} the distribution of \mbox{\boldmath $\sigma p$}
and \mbox{\boldmath $\sigma r$} strengths was found to be very close
to each other). Therefore:
\begin{equation}
\langle r\vert \sum_{i=1}^{A}(\mbox{\boldmath $\sigma r$})_{i}
\sum_{i=1}^{A}(\mbox{\boldmath $\sigma p$})_{i}\vert r\rangle=
i\omega m\langle r\vert \sum_{i=1}^{A}
(\mbox{\boldmath $\sigma r$})_{i}
\sum_{i=1}^{A}(\mbox{\boldmath $\sigma r$})_{i}\vert r\rangle \ .
\label{b3}
\end{equation}
The last matrix element in this equation is the total spin-dipole
strength and therefore equals to $N^{2}$. Eq.\ (\ref{b2}) is now:
\begin{equation}
\langle\tilde{\Phi}_{s}^{^{(-)}}\vert H_{PD}\vert d\rangle
\langle d\vert V^{PV}\vert r\rangle
=gi\omega m N \langle\tilde{\Phi}_{s}^{^{(-)}}\vert H_{PD}\vert d
\rangle \ .
\label{b4}
\end{equation}
Let us write the state $\vert r\rangle$ in the form:
\begin{equation}
\vert r\rangle=a_{r}\vert \phi_{p_{1/2}}^{B}\rangle+\vert f\rangle
\label{b5}
\end{equation}
where $\vert f\rangle$ stands for more complicated components. The
wave function of the spin-dipole one can write as:
\begin{equation}
\vert SD\rangle=\frac{1}{N}
\langle\phi_{s_{1/2}}^{B}\vert\mbox{\boldmath $\sigma r$}\vert
\phi_{p_{1/2}}^{B}\rangle a_{r}\vert\phi_{s_{1/2}}^{B}\rangle+\vert
f'\rangle \ ,
\label{b6}
\end{equation}
where $\vert f'\rangle$ again are complicated states. Because we take
into account only the one-body part of $H_{DP}$ only the
single-particle component in the spin-dipole doorway will contribute.
Thus:
\begin{eqnarray}
&&\langle\tilde{\Phi}_{s}^{^{(-)}}\vert H_{PD}\vert d\rangle
\langle d\vert V^{PV}\vert r\rangle \nonumber \\
&&=gi\omega m a_{r}\langle\tilde{\Phi}_{s}^{^{(-)}}\vert H_{PD}\vert
\phi_{s_{1/2}}^{B}\rangle \langle\phi_{s_{1/2}}^{B}\vert
\mbox{\boldmath $\sigma r$}\vert
\phi_{p_{1/2}}^{B}
\rangle \ .
\label{b7}
\end{eqnarray}
One can write for harmonic oscillator wave functions:
\begin{equation}
gi\omega m\langle\phi_{s_{1/2}}^{B}\vert\mbox{\boldmath $\sigma r$}
\vert\phi_{p_{1/2}}^{B}\rangle=
\langle\phi_{s_{1/2}}^{B}\vert V^{PV}\vert\phi_{p_{1/2}}^{B}\rangle
\label{b8}
\end{equation}
The last matrix element in Eq.\ (\ref{b7}) is just the single-particle
escape amplitude \mbox{$\gamma_{s}^{\uparrow}(E_{s}=E_{r})$} in
Eq.\ (\ref{51}). Using one-body potentials, only the single-particle
part \mbox{$a_{r}\vert \phi_{p_{1/2}}^{B}\rangle$} will contribute to
the matrix element
\mbox{$\langle r\vert H_{PR}\vert \Phi_{p}^{^{(+)}}\rangle$} in
Eq.\ (\ref{43}). The coefficients $a_{r}$ will drop out and we finally
obtain:
\begin{equation}
P_(E_{r})=-2\frac{Re(\gamma_{s}^{\uparrow}(E_{r})
\langle\phi_{s_{1/2}}^{B}\vert V^{PV}\vert\phi_{p_{1/2}}^{B}\rangle
\gamma_{p}^{\uparrow}(E_{r}))}
{\vert\gamma_{p}^{\uparrow}(E_{r})\vert^{2}\cdot(E_{r}-E_{d})} \ ,
\label{b9}
\end{equation}
where \mbox{$\gamma_{p}^{\uparrow}(E_{r})$} is the singe-particle
escape amplitude for the $p_{1/2}$ resonance, given in Eq.\
(\ref{51}).

\begin{figure}
\caption{A schematic representation of the doorway state approximation
in the calculation of the parity violating spreading width (see
text).}
\label{fig:gamma-d-pv}
\end{figure}

\begin{figure}
\caption{(a) The unprojected and projected $p_{1/2}$ continuum wave
functions for \mbox{$E_{n}=1$ eV}. (b)~The unprojected and projected
$s_{1/2}$ continuum wave functions for \mbox{$E_{n}=1$ eV}. The wave
functions are given in arbitrary units.}
\label{fig:rad-ort-no-wf}
\end{figure}

\begin{figure}
\caption{The calculated compound term of the asymmetry $P_{comp}$
using the doorway state approximation. The results are given as a
function of the neutron energy $E_{n}$. The two curves correspond to
calculations in which different shell model potentials were used.}
\label{fig:doorw-P(E)}
\end{figure}

\begin{figure}
\caption{The calculated direct term of the asymmetry $P$ calculated as
a function of the neutron energy $E_{n}$. The two curves correspond to
calculations in which different shell model potentials were used.}
\label{fig:direct-P(E)}
\end{figure}

\begin{figure}
\caption{(a) The integral in Eq.\ (\protect\ref{72}) calculated as a
function of the upper limit virtual energy $E'$ (arbitrary units).
(b)~A comparison between the exact and surface-peaked approximation.
The ratio of the product of the left hand side of
Eqs.\ (\protect\ref{70}) and (\protect\ref{71}) of the product of the
right hand sides of these equations is shown for the case of
\mbox{$E=1$ eV} as a function of the virtual energy $E'$.}
\label{fig:surf-peak-int}
\end{figure}

\begin{table}
\caption{The values $P(E_{r}=1eV)/\epsilon$, for different terms for
the case of orthogonalized continuum wave functions calculated for two
shell model potentials. Potential I has \mbox{$V_{0}=-69.7$ MeV},
potential II has \mbox{$V_{0}=-59.5$ MeV}.
\label{table1}}
\begin{tabular}{ldd}
 &\multicolumn{2}{d}{$P(E_{r}=1eV)/\epsilon$ in \%} \\
     term                & potential I & potential II \\  \tableline
compound (doorway)         & -0.95 & -0.44 \\
direct 		           &  0.16 &  0.12 \\
channel coupling resonance & -0.32 & -0.15 \\
 $\sum{P_{i}}$	           & -1.11 & -0.47 \\
\end{tabular}
\end{table}

\end{document}